\documentclass[epj,nopacs]{svjour}\pdfoutput=1
\usepackage{graphicx}%
\DeclareGraphicsExtensions{.pdf,.png,.jpg}%
\graphicspath{{./plots/}}%
\usepackage{paralist}%
\usepackage{multirow}%
\usepackage{pdflscape}%
\usepackage{rotating}%
\usepackage{color}%
\usepackage[usenames,dvipsnames,svgnames,table]{xcolor}%
\usepackage{xspace}%
\usepackage{booktabs}%
\usepackage{footmisc}%
\usepackage{ulem}%
\usepackage[mathlines]{lineno}%
\renewcommand{\arraystretch}{1.2} % make lines a bit larger for tables
\usepackage[british]{babel}%
\usepackage[utf8]{inputenc}
\usepackage{afterpage}
\usepackage{amsmath}%
\usepackage{hyperref}%

\newcommand{\TODOE}[1]{}
\newcommand{\TODOF}[1]{}

\newcommand{\hftwo}{\hspace*{\fill}}

\providecommand{\ie}{\mbox{i.e.}\xspace}     %i.e.
\providecommand{\eg}{\mbox{e.g.}\xspace}     %e.g.
\newcommand{\ep}{\ensuremath{\text{ep}}\xspace}

\newcommand{\HERA}{HERA\xspace}

\newcommand{\LHC}{LHC\xspace}

\newcommand{\pp}{\ensuremath{\text{pp}}\xspace}
\newcommand{\ppbar}{\ensuremath{\text{p}\bar{\text{p}}}\xspace}

\newcommand{\SPPS}{Sp$\bar{\text{p}}$S\xspace}
\newcommand{\Tevatron}{Tevatron\xspace}
\newcommand{\RHIC}{RHIC\xspace}

\newcommand{\fastjet}{{\textrm{FastJet}}\xspace}
\newcommand{\fastNLO}{{\textrm{fastNLO}}\xspace}

\newcommand{\LHAPDF}{{\textrm{LHAPDF}}\xspace}

\newcommand{\NLOJETPP}{{\textrm{NLOJet++}}\xspace}

\newcommand{\CRunDec}{{\textrm{CRunDec}}\xspace}

\newcommand{\our}{\text{common}\xspace} % "o" stands for "our" :-)
\newcommand{\GeV}{\ensuremath{\,\text{Ge\hspace{-.08em}V}}\xspace}
\newcommand{\GeVsq}{\ensuremath{\,\text{Ge\hspace{-.08em}V}^2}\xspace}
\newcommand{\TeV}{\ensuremath{\,\text{Te\hspace{-.08em}V}}\xspace}

\newcommand{\pbinv}{\mbox{\ensuremath{\,\text{pb}^\text{$-$1}}}\xspace}
\newcommand{\fbinv}{\mbox{\ensuremath{\,\text{fb}^\text{$-$1}}}\xspace}

\newcommand{\as}{\ensuremath{\alpha_{\text{s}}}\xspace}

\newcommand{\asmu}{\ensuremath{\as(\mur)}\xspace}
\newcommand{\asmz}{\ensuremath{\as(M_{\text{Z}})}\xspace}
\newcommand{\asmzpdf}{\ensuremath{\alpha_\text{s}^\text{PDF}(M_{\text{Z}})}\xspace}

\newcommand{\chisq}{\ensuremath{\chi^2}\xspace}
\newcommand{\ndf}{\ensuremath{n_{\text{dof}}}\xspace}
\newcommand{\chisqndf}{\ensuremath{\chi^2/\ndf}\xspace}

\newcommand{\kt}{\ensuremath{k_\text{T}}\xspace}

\newcommand{\mur}{\ensuremath{\mu_{\text{r}}}\xspace}
\newcommand{\muf}{\ensuremath{\mu_{\text{f}}}\xspace}

\newcommand{\order}[1]{\ensuremath{\mathcal{O}\left(#1\right)}\xspace}
\newcommand{\et}{\ensuremath{E_{\text{T}}}\xspace}
\newcommand{\pdfas}{\mbox{PDF\as}\xspace}
\newcommand{\pdfset}{\mbox{PDFset}\xspace}

\newcommand{\pt}{\ensuremath{p_{\text{T}}}\xspace}

\newcommand{\ptjet}{\ensuremath{p_{\text{T}}^{\rm jet}}\xspace}

\newcommand{\ptmax}{\ensuremath{p_\text{T}^\text{max}}\xspace}

\newcommand{\Qsq}{\ensuremath{Q^2}\xspace}

\newcommand{\yDIS}{\ensuremath{y_{\rm  DIS}}\xspace}

\newcommand{\MSbar}{\ensuremath{\overline{\text{MS}}}\xspace}

\title{Determination of the strong coupling constant from inclusive
  jet cross section data from multiple experiments}
\titlerunning{Determination of the strong coupling constant from
  multiple experiments}

\author{Daniel Britzger\inst{1}\and Klaus Rabbertz\inst{2}\and Daniel
  Savoiu\inst{2}\and Georg Sieber\inst{2}\and Markus Wobisch\inst{3}}
\authorrunning{D.~Britzger et al.}

\institute{Physikalisches Institut, Universität Heidelberg, Im
  Neuenheimer Feld 226, 69120 Heidelberg, Germany,
  \email{britzger@physi.uni-heidelberg.de}\and Institut für
  Experimentelle Teilchenphysik, KIT, Postfach 6980, 76128 Karlsruhe,
  Germany, \email{klaus.rabbertz@cern.ch, georg.sieber@cern.ch,
    daniel.savoiu@cern.ch}\and Department of Physics, Louisiana Tech
  University, 600 Dan Reneau Dr., Ruston, LA, USA,
  \email{wobisch@latech.edu}}

\abstract{Inclusive jet cross section measurements from the ATLAS,
  CDF, CMS, D0, H1, STAR, and ZEUS experiments are explored for
  determinations of the strong coupling constant \asmz.  Various jet
  cross section data sets are reviewed, their consistency is examined,
  and the benefit of their simultaneous inclusion in the \asmz
  determination is demonstrated.  Different methods for the
  statistical analysis of these data are compared and one method is
  proposed for a coherent treatment of all data sets.  While the
  presented studies are based on next-to-leading order in perturbative
  quantum chromodynamics (pQCD), they lay the groundwork for
  determinations of \asmz at next-to-next-to-leading order. %
  \keywords{Jets, QCD Phenomenology, strong coupling constant}%
}

\begin{document}
\maketitle
\flushbottom

\section{Introduction}
\label{sec:intro}

The strong coupling constant, \as, is one of the least precisely known
fundamental parameters in the Standard\linebreak Model of particle physics.
Because of its importance for precision phenomenology at the \LHC and
elsewhere, large efforts have been undertaken in the past decades to
reduce uncertainties in determinations of
\as~\cite{Patrignani:2016xmw,Dissertori:2015tfa,d'Enterria:2015toz,ISBN:9783319421155}.

With the advent of modern particle detectors and sophisticated
algorithms for their simulation and calibration, jet measurements have
become very precise.  Many deter\-minations of \as in deep-inelastic
scattering (DIS) and in hadron-hadron collisions are therefore based
on measurements of the inclusive jet cross section, which is directly
proportional to \as in DIS in the Breit frame and $\as^2$ in
hadron-hadron collisions.  Using the most precise predictions of
perturbative quantum chromodynamics (pQCD) available at the time, all
previous \as extractions (except for ref.~\cite{Andreev:2017vxu}) were
performed at next-to-leading order (NLO) in \as.  Their total
uncertainty is dominated by the contribution related to the
renormalisation scale dependence of the NLO pQCD results.  The recent
completion of next-to-next-to-leading order (NNLO) predictions for the
inclusive jet cross section~\cite{Currie:2016bfm,Currie:2017tpe}
promises a considerable reduction of the renormalisation scale
dependence and will allow the inclusion of \as results from inclusive
jet data in future determinations of the world average value of
\as~\cite{Patrignani:2016xmw}.

A determination of \as at NNLO from jet measurements in hadron-hadron
collisions is still not readily achievable, because the new NNLO pQCD
calculations are computationally very demanding and cannot yet be
repeated quickly for different parton distribution functions (PDFs) or
values of \asmz.  In preparation of such a determination, it is
desirable to study a simultaneous analysis of data sets from different
processes and experiments.  This study includes an investigation of
the consistency of the various data sets and an estimation of the
reduction of the experimental contributions to the \as uncertainty.
The groundwork for these two aspects is presented in this article.

We review inclusive jet cross section data over a wide kinematic
range, from different experiments for various initial states and
centre-of-mass energies, and study their potential for determinations
of \as.  The consistency of the diverse data sets is examined and the
benefit of their simultaneous inclusion is demonstrated. Different
methods for the statistical analysis of the data are compared and one
method is proposed for a coherent treatment of all data sets in an
extraction of \asmz.

The article is structured as follows: The experimental data sets and
the theoretical predictions are introduced in sections~\ref{sec:data}
and~\ref{sec:theory}, respectively.  Methods and results from previous
\as determinations by different experimental collaborations are
discussed and employed in section~\ref{sec:methods}. The strategy for
a determination of \as from multiple data sets and the final result
are presented in section~\ref{sec:common}.

% -----------------------------------------------------------
% -----------------------------------------------------------
% -----------------------------------------------------------
\section{Experimental data}
\label{sec:data}

The first measurement of the inclusive jet cross section has been
performed in 1982 by the UA2 Collaboration at the \SPPS collider at a
centre-of-mass energy of 540\GeV~\cite{Banner:1982kt}.  Further
measurements have been conducted at centre-of-mass energies of

\begin{compactitem}
\item 540\GeV, 546\GeV, and 630\GeV at the \ppbar collider \SPPS by
  the UA1~\cite{Arnison:1983gw,Arnison:1986vk} and UA2
  experiments~\cite{Appel:1985rm},
\item 546\GeV, 630\GeV, 1.8\TeV, and 1.96\TeV at the \Tevatron \ppbar
  collider by the CDF~\cite{Abe:1992bk,Affolder:2001fa,Abulencia:2005yg,Abulencia:2005jw,Abulencia:2007ez,Aaltonen:2008eq} and D0 experiments~\cite{Abbott:2000ew,Abbott:2000kp,Abazov:2008ae},
\item 300\GeV and 320\GeV at the \ep collider \HERA by the
  H1~\cite{Adloff:2000tq,Adloff:2002ew,Adloff:2003nr,Aktas:2007aa,Aaron:2009vs,Aaron:2010ac,Andreev:2014wwa,Andreev:2016tgi} and
  ZEUS experiments \cite{Breitweg:1998ur,Chekanov:2002be,Chekanov:2002ru,Chekanov:2006xr,Chekanov:2006yc,Abramowicz:2010ke,Abramowicz:2012jz},
\item 200\GeV in \pp collisions at \RHIC by the STAR
  experiment~\cite{Abelev:2006uq},
\item and of 2.76\TeV, 7\TeV, 8\TeV, and 13\TeV in \pp collisions at
  the \LHC by the ALICE~\cite{Abelev:2013fn}, ATLAS~\cite{Aad:2010ad,Aad:2011fc,Aad:2013lpa,Aad:2014vwa,Aaboud:2017dvo,Aaboud:2017jcu}, and CMS
  experiments~\cite{Chatrchyan:2011ab,Chatrchyan:2012gw,Chatrchyan:2012bja,Chatrchyan:2014gia,Khachatryan:2015luy,Khachatryan:2016wdh,Khachatryan:2016mlc}.
\end{compactitem}

\noindent{}While earlier measurements established the inclusive jet
cross section as a useful quantity to study QCD, large experimental
uncertainties limited their use for QCD phenomenology.  When the NLO
pQCD corrections were computed~\cite{Ellis:1990ek,Ellis:1992en,Giele:1994gf}, studies revealed collinear- or
infrared-safety issues in the jet definitions used in the experimental
measurements~\cite{Seymour:1997kj}.  These issues were subsequently
addressed and improved jet definitions were
developed~\cite{Blazey:2000qt,Salam:2009jx} and applied in recent
measurements.

Previously, \asmz determinations were based on inclusive jet cross
section data from individual experiments, as summarised in
table~\ref{tab:previousasmz}.  An extraction of \asmz from multiple
inclusive jet cross section data sets has not been performed so far,
except in the context of global PDF analyses, in which PDFs and \asmz
are determined simultaneously.  These analyses, however, require data
for a variety of measured quantities~\cite{Adloff:2000tq,Chekanov:2005nn,Abramowicz:2015mha,Ball:2011us,Harland-Lang:2014zoa}.  In this article, \asmz is
determined in a fit to multiple inclusive jet cross section
measurements from experiments at \HERA, \RHIC, the \Tevatron, and the
\LHC.  The analysis is based on one selected measurement from each,
the H1, ZEUS, STAR, CDF, D0, ATLAS, and CMS collaborations, as listed
in table~\ref{tab:datasets}.

% ----- Previously extracted alpha_s values from inclusive jets
\begin{table*}[tbp]
%  \footnotesize
  \begin{center}
    \begin{tabular}{l@{\hskip4pt}l@{\hskip4pt}ll@{\hskip2pt}l@{\hskip0pt}l}
      \toprule
      {\bf Publication} & {\bf data} & {\bf comment} &
      \multicolumn{3}{c}{\bf\boldmath \asmz}\\\midrule
      H1~\cite{Andreev:2014wwa} & H1~\cite{Andreev:2014wwa}
      & \HERA II, high \Qsq
      & $0.1174$ & $(22)_\text{exp}$ & $(50)_\text{theo}$ \\
      D0~\cite{Abazov:2009nc} & D0~\cite{Abazov:2008ae}
      & aNNLO, 22 points
      & $0.1161$ & $(^{+34}_{-33})_\text{exp}$ & $(^{+29}_{-35})_\text{theo}$\\
      D0~\cite{Abazov:2009nc} & D0~\cite{Abazov:2008ae}
      & NLO, 22 points
      & $0.1202$ & $(^{+72}_{-59})_\text{tot}$ & \\
      CMS~\cite{Khachatryan:2014waa} & CMS~\cite{Chatrchyan:2012bja}
      & 7\TeV, 5.0\fbinv
      & $0.1185$ & $(19)_\text{exp}$ & $(^{+60}_{-37})_\text{theo}$\\\midrule
      H1~\cite{Adloff:2000tq} & H1~\cite{Adloff:2000tq}
      & \HERA I, $\sqrt{s} = 300\GeV$
      & $0.1186$ & $(30)_\text{exp}$ & $(51)_\text{theo}$ \\
      H1~\cite{Aktas:2007aa} & H1~\cite{Aktas:2007aa}
      & \HERA I, $\sqrt{s} = 320\GeV$
      & $0.1193$ & $(14)_\text{exp}$ & $(^{+50}_{-34})_\text{theo}$ \\
      H1~\cite{Aaron:2010ac} & H1~\cite{Aaron:2010ac}
      & \HERA I, $\sqrt{s} = 320\GeV$, low \Qsq
      & $0.1180$ & $(18)_\text{exp}$ & $(^{+124}_{-93})_\text{theo}$ \\
      H1~\cite{Andreev:2017vxu} & H1~\cite{Adloff:2000tq,Aktas:2007aa,Aaron:2010ac,Andreev:2014wwa,Andreev:2016tgi}
      & \HERA I+II, NNLO
      & $0.1152$ & $(20)_\text{exp}$ & $(27)_\text{theo}$ \\
      ZEUS~\cite{Chekanov:2002be} & ZEUS~\cite{Chekanov:2002be}
      & $\sqrt{s} = 300\GeV$, $\Qsq>500\GeVsq$
      & $0.1212$ & $(^{+29}_{-35})_\text{exp}$ & $(^{+28}_{-27})_\text{theo}$ \\
      ZEUS~\cite{Chekanov:2006yc} & ZEUS~\cite{Chekanov:2006xr}
      & $d\sigma/d\Qsq$, $\Qsq>500\GeVsq$
      & $0.1207$ & $(^{+38}_{-36})_\text{exp}$ & $(^{+22}_{-23})_\text{theo}$ \\
      CDF~\cite{Affolder:2001hn} & CDF~\cite{Affolder:2001hn}
      & 1.8\TeV, 87\pbinv
      & $0.1178$ & $(^{+81}_{-95})_\text{exp}$ & $(^{+92}_{-75})_\text{theo}$ \\
      CMS~\cite{Khachatryan:2016mlc} & CMS~\cite{Khachatryan:2016mlc}
      & 8\TeV, 19.7\fbinv
      & $0.1164$ & $(^{+14}_{-15})_\text{exp}$ & $(^{+59}_{-40})_\text{theo}$\\\midrule
      W.T.~Giele et al.~\cite{Giele:1995kb} & CDF~\cite{Abe:1991ea}
      & 1.8\TeV, 4.2\pbinv
      & $0.121$ & $(8)_\text{exp}$ & $(5)_\text{theo}$ \\
      B.~Malaescu et al.~\cite{Malaescu:2012ts} & ATLAS~\cite{Aad:2011fc}
      & 7\TeV, 37\pbinv
      & $0.1151$ & $(47)_\text{exp}$ & $(^{+51}_{-40})_\text{theo}$ \\
      T.~Biekötter et al.~\cite{Biekotter:2015nra} & H1~\cite{Andreev:2014wwa}
      & aNNLO
      & $0.122$ & $(2)_\text{exp}$ & $(13)_\text{theo}$ \\
      T.~Biekötter et al.~\cite{Biekotter:2015nra} & H1~\cite{Andreev:2014wwa}
      & NLO
      & $0.115$ & $(2)_\text{exp}$ & $(5)_\text{theo}$\\\bottomrule
    \end{tabular}
    \caption{Summary of previous determinations of \asmz from
      inclusive jet cross sections. The upper part lists the recent
      \asmz extractions from double-differential inclusive jet cross
      sections by experimental collaborations that are studied in more
      detail in this work. The middle and lower parts summarise
      further determinations of \asmz by experimental collaborations
      and by independent authors, respectively.  The results in
      refs.~\cite{Abazov:2009nc} and~\cite{Biekotter:2015nra} are
      reported for approximate NNLO (aNNLO) and NLO used for the pQCD
      predictions. In ref.~\cite{Abazov:2009nc} only 22 out of the 110
      D0 data points were used in the \asmz extraction; the
      decomposition of the uncertainties is only provided for the
      aNNLO result.  In case of ref.~\cite{Malaescu:2012ts}, we only
      consider scale, PDF, and NP related uncertainties as theoretical
      uncertainty for reasons of comparability to the other listed
      results.}
    \label{tab:previousasmz}
  \end{center}
\end{table*}

% ----- Overview of the inclusive jet data sets
\begin{table*}[tbp]
%  \scriptsize
  \renewcommand{\arraystretch}{1.3}
  \begin{center}
    \begin{tabular}{l@{\hskip4pt}c@{\hskip6pt}c@{\hskip4pt}c@{\hskip2pt}cccc@{\hskip0pt}}
      \toprule
      {\bf Data} & {\bf proc} & {\boldmath$\sqrt{s}$} & {\boldmath{$\mathcal{L}$}}
      & {\bf no.\ of} & {\bf jet algorithm} & {\bf
        {\boldmath$\pt,\,\et$}-range} & {\bf other kinematic} \\
      & & {\bf[TeV]} & {\boldmath{$[{\rm fb}^{-1}]$}} & {\bf points} & &
      {\bf [GeV]} & {\bf ranges}\\
      \midrule
      & & & & & & & $150<\Qsq<15\,000\GeVsq$ \\
      H1~\cite{Andreev:2014wwa} & \ep & 0.32 & 0.35 & 24 &
      \kt, $R=1.0$ & $7<\pt<50$ & $0.2<\yDIS<0.7$ \\
      & & & & & & & $-1.0<\eta<2.5$\\
      \midrule
      & & & & & & & $\Qsq>125\GeVsq$ \\
      ZEUS~\cite{Chekanov:2006xr} & \ep & 0.32 & 0.082 & 30 &
      \kt, $R=1.0$ & $\et>8$ & $|\cos\gamma_h|<0.65$ \\
      & & & & & & & $-2.0<\eta<1.5$\\
      \midrule
      STAR~\cite{Abelev:2006uq} & \pp & 0.20 & 0.0003 & 9 &
      MP, $R=0.4$ &  $7.6 < \pt <48.7$ & $0.2<|\eta|<0.8$\\
      \midrule
      CDF~\cite{Abulencia:2007ez} & \ppbar & 1.96 & 1.0 & 76 &
      \kt, $R=0.7$ & $54<\pt<527$ & $|y|<2.1$\\
      \midrule
      D0~\cite{Abazov:2008ae} & \ppbar & 1.96 & 0.7 & 110 &
      MP, $R=0.7$ & $50<\pt<665$ & $|y|<2.0$\\
      \midrule
      ATLAS~\cite{Aad:2014vwa} & \pp & 7.0 & 4.5 & 140 &
      anti-\kt, $R=0.6$ & $100<\pt<1992$ & $|y|<3.0$\\
      \midrule
      CMS~\cite{Khachatryan:2014waa} & \pp & 7.0 & 5.0 & 133 &
      anti-\kt, $R=0.7$ & $114<\pt<2116$ & $|y|<3.0$\\
      \bottomrule
    \end{tabular}
  \end{center}
  \caption{Overview of the inclusive jet data sets used in the \as
    determinations.  For each data set the process (proc), the
    centre-of-mass energy $\sqrt{s}$, the integrated luminosity
    $\mathcal{L}$, the number of data points, and the jet algorithm are
    listed.  In case of \ep collider data, the kinematic range may be
    defined by the four-momentum transfer squared \Qsq, the inelasticity
    \yDIS, or the angle of the hadronic final state $|\cos\gamma_h|$ of
    the NC DIS process.  In all cases, jets are required to be within a
    given range of pseudorapidity $\eta$ or rapidity $y$ in the laboratory
    frame.}
  \label{tab:datasets}
\end{table*}

Whenever experiments provide multiple measurements, we include those
measured with a collinear- and infrared-safe jet algorithm
(\kt~\cite{Ellis:1993tq} or anti-\kt~\cite{Cacciari:2008gp}) and with
a larger jet size parameter $R$, which improves the stability of
fixed-order pQCD calculations.  The STAR experiment published
inclusive jet data collected by two different triggers with partially
overlapping jet \pt ranges.  We choose the data set collected with the
trigger covering the higher jet \pt range from 7.6\GeV up to 50\GeV.
The measurements from the STAR and D0 experiments are using the
midpoint cone jet algorithms (MP)~\cite{Blazey:2000qt}.  The
infrared-unsafety of this jet algorithm~\cite{Seymour:1997kj}
prohibits NNLO pQCD predictions for these data sets, but it does not
affect calculations at NLO\@.  Four new
measurements~\cite{Andreev:2016tgi,Khachatryan:2016mlc,Aaboud:2017dvo,Aaboud:2017jcu} could not be included in this study;
they are left for a future extension.

% -----------------------------------------------------------
% -----------------------------------------------------------
% -----------------------------------------------------------
\section{Theoretical predictions and tools}
\label{sec:theory}

Predictions for the inclusive jet cross section in processes with
initial-state hadrons are calculated as the convolution of the
partonic cross section $\hat{\sigma}$ (computed in pQCD) and the PDFs
of the hadron(s).  The inclusive jet cross section in hadron-hadron
collisions can be written
as~\cite{ISBN:9780521581899,Patrignani:2016xmw}
\begin{multline}
  \sigma_\text{pQCD,hh}(\mur,\muf) =
  \sum_{i,j} \int dx_1 \int dx_2 \:\\
  f_{i/h_1}(x_1,\muf)\,f_{j/h_2}(x_2,\muf) \:
  \hat{\sigma}_{ij\rightarrow\text{jet}+X}(\mur,\muf) \: ,
  \label{eq:qcdcs}
\end{multline}
where the sum is over all combinations of parton flavors $i$ and $j$
(quarks, anti-quarks, and the gluon).  The $f_{i,j/h_{1,2}}$ denote
the PDFs for the parton flavours $i$ or $j$ in the initial-state
hadrons $h_1$ and $h_2$, and $x_{1}$ and $x_{2}$ correspond to the
fractional hadron momenta carried by the partons $i$ and $j$,
respectively.  The partonic cross section
$\hat{\sigma}_{ij\rightarrow\text{jet}+X}$ is computed as a
perturbative expansion in \as as
\begin{equation}
  \hat{\sigma}_{ij\rightarrow\text{jet}+X}(\mur,\muf) \: = \:
  \sum_{n}  \alpha^n_s(\mur) \:
  c^{(n)}_{ij\rightarrow\text{jet}+X}(\mur,\muf) \: ,
\end{equation}
where the $c^{(n)}_{ij\rightarrow\text{jet}+X}$ are computed from the
pQCD matrix elements and the sum is over all orders of \as taken into
account in the perturbative calculation.  The renormalisation and
factorisation scales are labelled $\mur$ and $\muf$, respectively.
For inclusive jet production in hadron-hadron collisions, the first
non-vanishing order (\ie the leading order, LO) is given by $n=2$,
while $n=3$ corresponds to the NLO corrections.  For inclusive jet
production in DIS in the Breit frame the partonic cross sections are
convoluted with a single PDF and the LO (NLO) contribution is given by
$n=1$ ($n=2$).  Hence, inclusive jet production in \pp, \ppbar, and
\ep collisions is sensitive to \as already at LO\@.

For transverse jet momenta at the \TeV scale accessible at the \LHC,
electroweak (EW) tree-level effects of \order{\alpha\as,\alpha^2} and
loop effects of \order{\alpha\as^2} become
sizeable~\cite{Dittmaier:2012kx}.  A recent study of the complete set
of QCD and EW NLO corrections has been presented in
ref.~\cite{Frederix:2016ost}.

Non-perturbative (NP) corrections to the cross section due to
multiparton interactions and hadronisation can be estimated by using
Monte Carlo (MC) event generators.  An overview of MC event generators
for the \LHC is presented in ref.~\cite{Buckley:2011ms}.  The size of
this correction depends on the jet size $R$, shrinks with increasing
jet \pt, and becomes negligible at the \TeV scale. The total theory
prediction for the inclusive jet cross section is given by
\begin{equation}
  \sigma_\text{theory} = \sigma_\text{pQCD}\cdot\,c_\text{EW}\cdot\,c_\text{NP}\,,
  \label{eq:cstheo}
\end{equation}
where $c_\text{EW}$ and $c_\text{NP}$ are the correction factors for
electroweak and non-perturbative corrections, respectively.

The partonic cross section is computed at NLO accuracy for five
massless quark flavours using the \NLOJETPP program
version~4.1.3~\cite{Nagy:2001fj,Nagy:2003tz} within the \fastNLO
framework at version~2~\cite{Kluge:2006xs,Britzger:2012bs} to allow
us fast recalculations for varying PDFs, scales \mur and \muf, and
assumptions on \asmz.  Jet algorithms are taken either from the
\fastjet software library~\cite{Cacciari:2011ma} or, for jet cross
sections in DIS, from \NLOJETPP\@.  The PDFs are evaluated via the
\LHAPDF interface~\cite{Whalley:2005nh,Buckley:2014ana} at version~6.
The running of \asmu is performed at 2-loop order using the package
\CRunDec with five massless quark
flavours~\cite{Chetyrkin:2000yt,Schmidt:2012az}.  The minimal
subtraction (\MSbar)
scheme~\cite{'tHooft:1972fi,'tHooft:1973mm,Bardeen:1978yd} has been
adopted for the renormalisation procedure in these calculations.

For the computation of the inclusive jet cross section in
hadron-hadron collisions, the renormalisation and factorisation
scales, \mur and \muf, are identified with each jet's \pt, \ie $\mur =
\muf = \ptjet$.  In neutral current (NC) DIS, the scales are chosen to
be ${\mur}^2 =\frac{1}{2}\left(Q^2 + ({\ptjet})^2\right)$ and
${\muf}^2 = Q^2$ as used by the H1
Collaboration~\cite{Andreev:2014wwa}.  Alternative scale choices have
been discussed with respect to NNLO predictions~\cite{Currie:2017tpe,Currie:2017ctp,Currie:2017tfd,Andreev:2017vxu}, but are beyond the
scope of this article.

The EW corrections, $c_\text{EW}$, relevant for the \LHC data are
provided by the experimental collaborations together with the data,
based on ref.~\cite{Dittmaier:2012kx}.  These are considered to have
negligible uncertainties.  Due to restrictions of the scale choices in
this calculation, the leading jet's transverse momentum, \ptmax, is
used to define the scales \mur and \muf.  The NP correction factors
$c_\text{NP}$, except for the STAR data~\cite{starnp}, are also
provided by the experimental collaborations, together with an estimate
of the corresponding uncertainty~\cite{Andreev:2014wwa,Chekanov:2006xr,Chekanov:2006yc,Abulencia:2007ez,Abazov:2008ae,Abazov:2009nc,Aad:2014vwa,Chatrchyan:2012bja,Khachatryan:2014waa}.

% -----------------------------------------------------------
% -----------------------------------------------------------
% -----------------------------------------------------------
\section{Comparison of three extraction methods for
  \texorpdfstring{\boldmath{\asmz}}{alpha\_s(M\_Z)}}
\label{sec:methods}

Commonly, the value of \asmz is determined from inclusive jet cross
sections in a comparison of pQCD predictions to the measurements.
These \asmz results therefore depend on details of the extraction
method such as the treatment of uncertainties in the characterisation
of differences between theory and data, or the evaluation and
propagation of theoretical uncertainties to the final result.  An
overview of previous determinations of \asmz from fits to inclusive
jet cross section data is provided in table~\ref{tab:previousasmz}.
We choose the three \asmz determinations performed by the
CMS~\cite{Khachatryan:2014waa}, D0~\cite{Abazov:2009nc}, and
H1~\cite{Andreev:2014wwa} collaborations listed in the upper part of
table~\ref{tab:previousasmz} for further study.

\noindent{}The three extraction methods differ in the following
aspects:
\begin{compactitem}
\item the definition of the \chisq function to quantify the
  agreement between theory and data,
\item the uncertainties considered in the \chisq function,
\item the strategy to determine the central result for \asmz,
\item the propagation of the uncertainties to the value of \asmz,
\item the choice of PDF sets,
\item the consideration of the \asmz dependence of the PDFs, and
\item the treatment of further theoretical uncertainties.
\end{compactitem}
To study the impact of these differences, we have implemented the
three methods in our computational framework and will refer to them as
``CMS-type'', ``D0-type'', and ``H1-type'', respectively.  Each method is
employed to extract \asmz from each of the individual data sets
selected in section~\ref{sec:data}, cf.\ also
table~\ref{tab:datasets}.  The experimental uncertainties and their
correlations are treated according to the respective prescriptions by
the experiments.  The CMS result was obtained with the CT10 PDF
set~\cite{Lai:2010vv}, and the D0 and H1 results with MSTW2008
PDFs~\cite{Martin:2009iq}.  The CMS-type and D0-type methods use the
entire \asmzpdf series available for the PDF set, whereas the H1-type
method uses a PDF determined with a value of $\asmzpdf=0.1180$.  The
resulting \asmz values are listed in table~\ref{tab:asvalues}.

% ----- table with all alpha_s reproduction fits
\begin{table*}[tbp]
%  \scriptsize
  \begin{center}
    \begin{tabular}{l|c@{\hskip4pt}cc@{\hskip4pt}cc@{\hskip4pt}c}
      \toprule
      {Fit method} %
      & \multicolumn{2}{c}{\bf CMS-type} %
      & \multicolumn{2}{c}{\bf D0-type} %
      & \multicolumn{2}{c}{\bf H1-type} %
      \\
      {PDF set} %
      & {MSTW2008} %
      & {CT10} %
      & {MSTW2008} %
      & {CT10} %
      & {MSTW2008} %
      & {CT10} %
      \\
      \midrule
      {\bf Data} %
      & \multicolumn{6}{c}{\bf\boldmath \asmz values with
        experimental uncertainties} %
      \\
      H1 %
      & $0.1172\,(28)$
      & $0.1172\,(28)$
      & $0.1161\,(27)$
      & $0.1164\,(26)$
      & \uline{$0.1174\,(22)$} %
      & \uline{$0.1180$}$\,(22)$ %
      \\
      ZEUS %
      & $0.1213\,(28)$
      & $0.1223\,(29)$
      & $0.1210\,(^{+28}_{-29})$
      & $0.1218\,(^{+30}_{-29})$
      & $0.1231\,(30)$
      & $0.1236\,(30)$
      \\
      STAR %
      & --- %
      & $0.1193\,(68)$
      & --- %
      & $0.1205\,(^{+54}_{-111})$
      & $0.1159\,(116)$
      & $0.1280\,(111)$
      \\
      CDF %
      & $0.1217\,(17)$
      & $0.1265\,(27)$
      & $0.1202\,(^{+10}_{-27})$
      & $0.1162\,(^{+22}_{-20})$
      & $0.1217\,(35)$
      & $0.1265\,(37)$
      \\
      D0 (22 pts., NLO) %
      & $0.1226\,(32)$
      & $0.1237\,(36)$
      & \uline{$0.1203$}$\,(^{+40}_{-42})$ %
      & $0.1191\,(^{+38}_{-45})$
      & $0.1219\,(50)$
      & $0.1232\,(51)$
      \\
      ATLAS %
      & $0.1220\,(9)$
      & $0.1258\,(15)$
      & $0.1204\,(^{+14}_{-5})$
      & $0.1241\,(9)$
      & $0.1206\,(15)$
      & $0.1270\,(16)$
      \\
      CMS %
      & \uline{$0.1162\,(14)$} %
      & \uline{$0.1188\,(19)$} %
      & $0.1158\,(12)$
      & $0.1162\,(19)$
      & $0.1140\,(21)$
      & $0.1217\,(23)$
      \\
      \bottomrule
    \end{tabular}
    \caption{Values of \asmz with experimental uncertainties obtained
      using the three extraction methods CMS-type, D0-type, and H1-type
      together with the CT10 or MSTW2008 PDF set at NLO\@. The
      underlined values can be directly compared with the results
      published in refs.~\cite{Khachatryan:2014waa,Abazov:2009nc,Andreev:2014wwa}.  Some fits to the STAR data do not exhibit a
      local minimum, in which case no value is listed.}
    \label{tab:asvalues}
  \end{center}
\end{table*}

In a first step, these results are compared to the ones obtained by
the CMS~\cite{Khachatryan:2014waa}, D0~\cite{Abazov:2009nc}, and
H1~\cite{Andreev:2014wwa} collaborations as listed in
table~\ref{tab:previousasmz}.  All three central results are
reproduced, the H1 result exactly, and the CMS and D0 results within
$+0.0003$ and $+0.0001$.  Such small differences can easily be caused
already by using different versions of \LHAPDF (\eg changes from
version~5 to version~6).  The experimental uncertainties of the CMS
and H1 analyses are exactly reproduced.\footnote{For the D0 analysis,
  the decomposition of uncertainties has been published only for their
  central result based on approximate NNLO pQCD and hence a comparison
  of the experimental uncertainty on \asmz for the NLO result is not
  possible.}

\begin{figure*}[tbp]
  \begin{center}
    \includegraphics[width=0.50\textwidth]{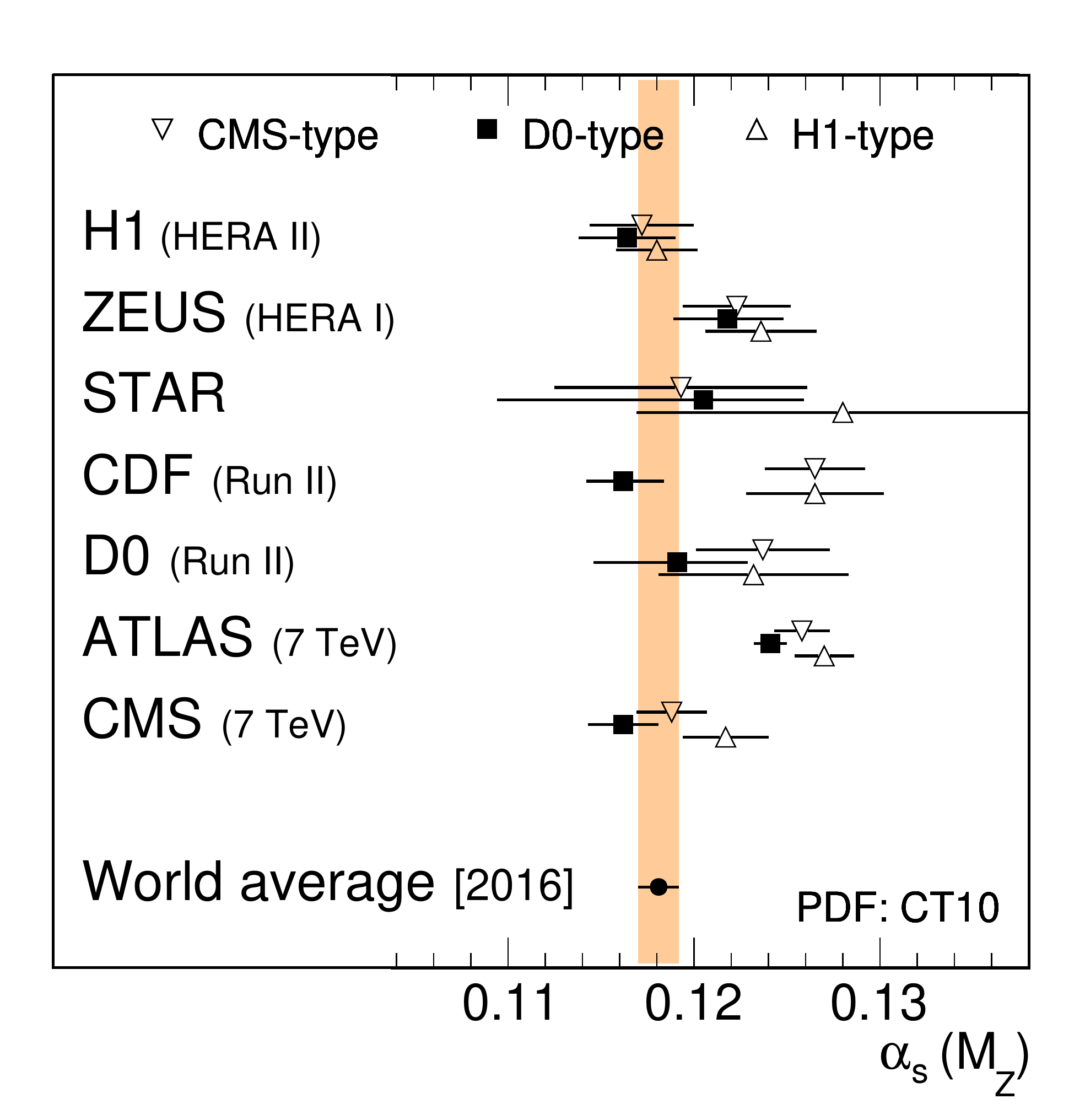}\hftwo%
    \includegraphics[width=0.50\textwidth]{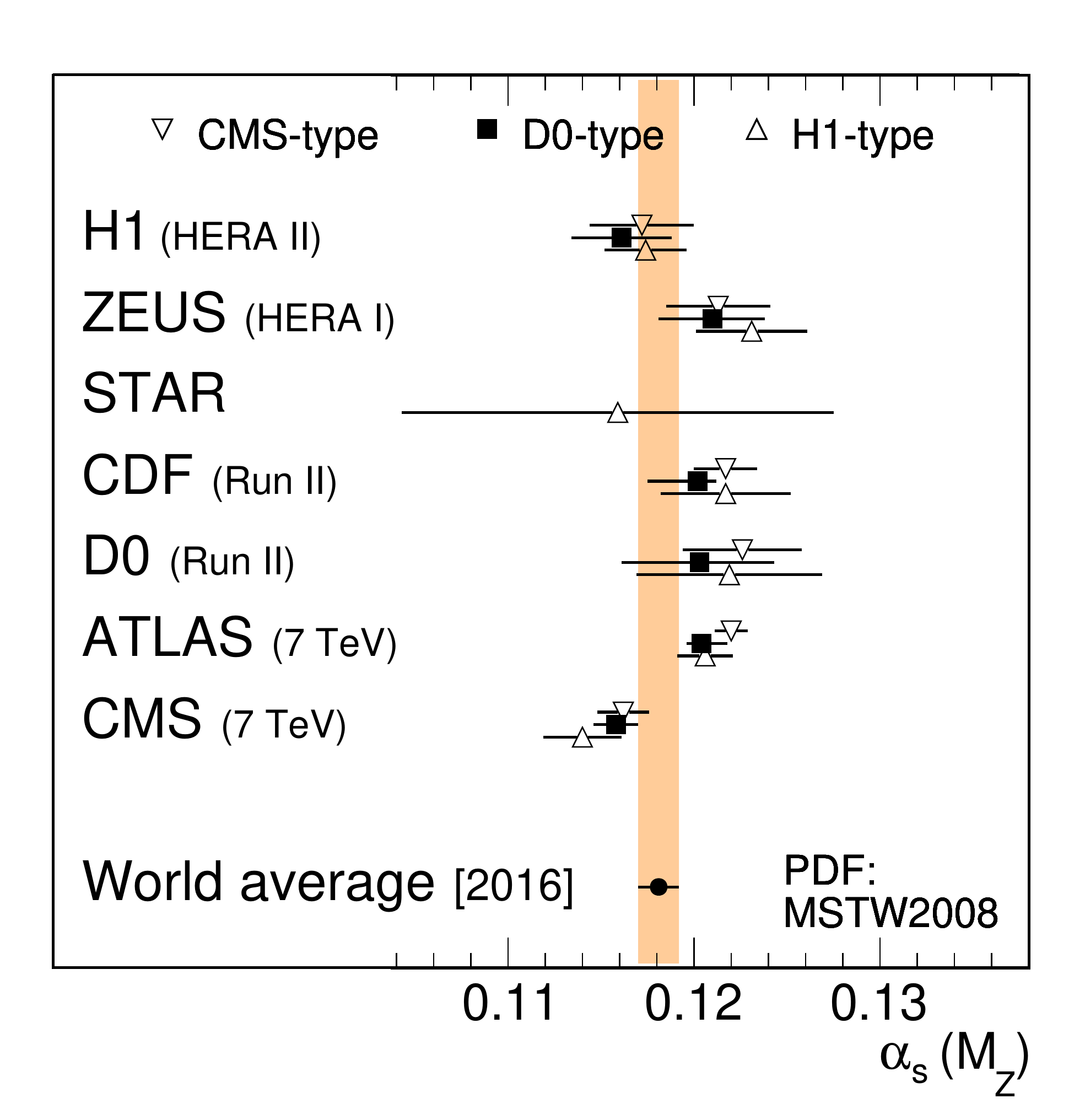}\\
    \includegraphics[width=0.50\textwidth]{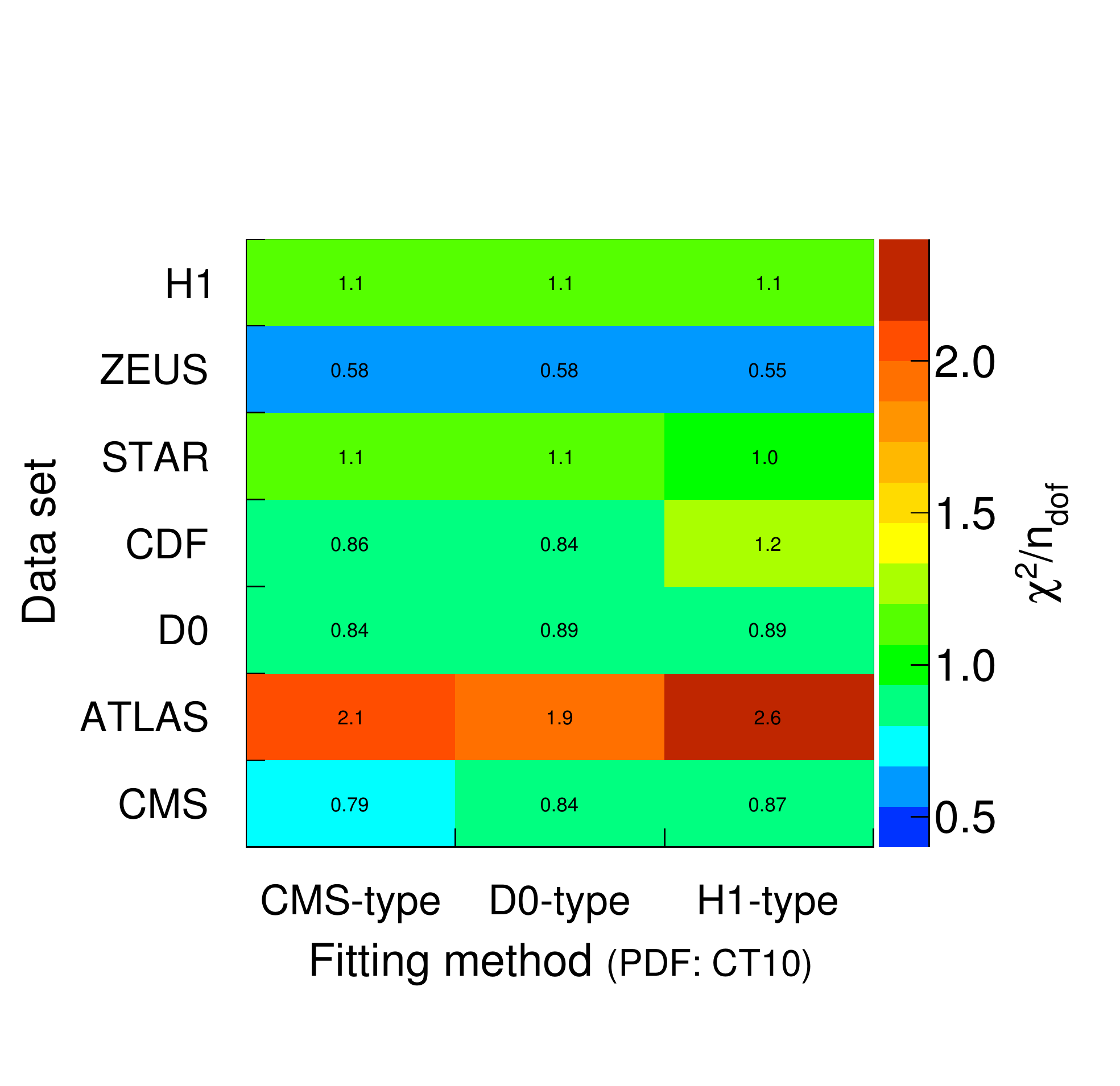}\hftwo%
    \includegraphics[width=0.50\textwidth]{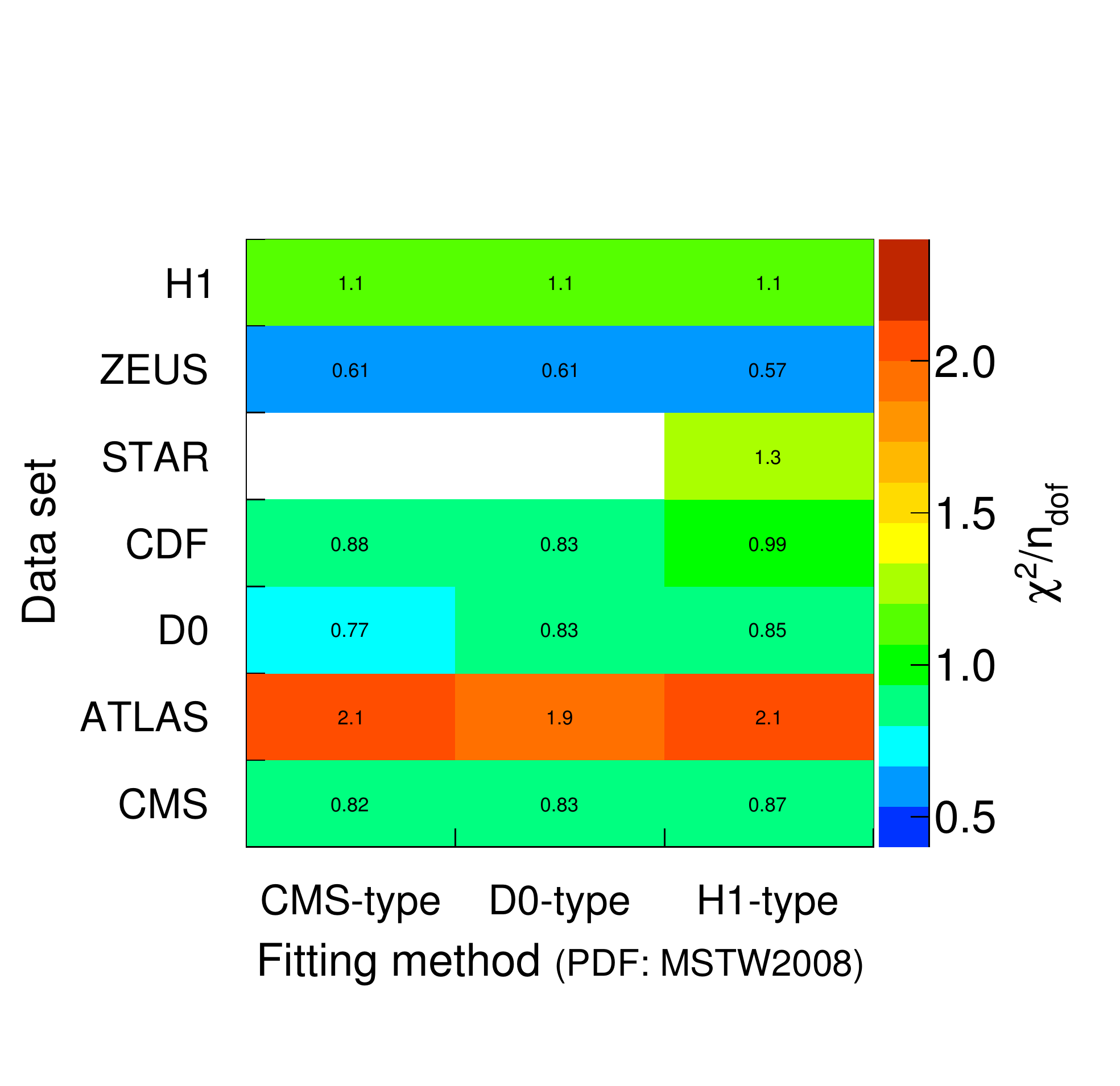}
  \end{center}
  \caption{Values of \asmz with experimental uncertainties obtained
    using the three extraction methods CMS-type, D0-type, and H1-type. The
    upper row compares the results for each data set employing the PDF
    set CT10 (left) or MSTW2008 (right). In addition, the world
    average value~\cite{Patrignani:2016xmw} is shown together with a
    band representing its uncertainty. The bottom row displays the
    values of \chisqndf for each fit using the CT10 (left) or MSTW2008
    PDF set (right). The colours illustrate the values of
    \chisqndf. For the STAR data and the MSTW2008 PDF set no local
    minimum was found in case of the CMS-type and D0-type fits (blank
    areas).}
  \label{fig:asvalues}
\end{figure*}

In a second step, the \asmz results and their experimental
uncertainties are compared to each other and their dependencies on the
extraction method and PDFs are studied.  The \asmz results determined
for each data set are displayed in figure~\ref{fig:asvalues} (top row)
for the three different extraction methods using CT10 PDFs (left) and
MSTW2008 PDFs (right).  For the STAR data, \asmz results cannot be
determined in case of the CMS-type and D0-type methods with MSTW2008
PDFs, since no local $\chisq$ minima are found.  In all other cases
the \asmz results obtained with MSTW2008 PDFs are rather independent
of the extraction method for all data sets.  This is different when
using CT10 PDFs: While in this case the extraction method has little
impact on the \asmz results from \HERA data (H1 and ZEUS), it notably
affects the results for the \LHC data (ATLAS and CMS), and has large
effects for the \Tevatron data (CDF and D0).  In the latter cases, the
D0-type method produces significantly lower \asmz results as compared
to the other two methods.

The \chisqndf values for the \asmz extractions are displayed in
figure~\ref{fig:asvalues} (bottom row) for the three extraction
methods using CT10 PDFs (left) and MSTW2008 PDFs (right).  Overall,
the fits exhibit reasonable values of\linebreak \chisqndf, thus indicating
agreement between theory and data.  Exceptions are observed for the
ZEUS data with rather low values of \chisqndf, and for the ATLAS data,
where the values of \chisqndf are large as also observed
elsewhere~\cite{Aaboud:2017dvo,Ball:2017nwa,Harland-Lang:2017ytb}.

\begin{figure}[tbp]
  \begin{center}
    \includegraphics[width=0.48\textwidth]{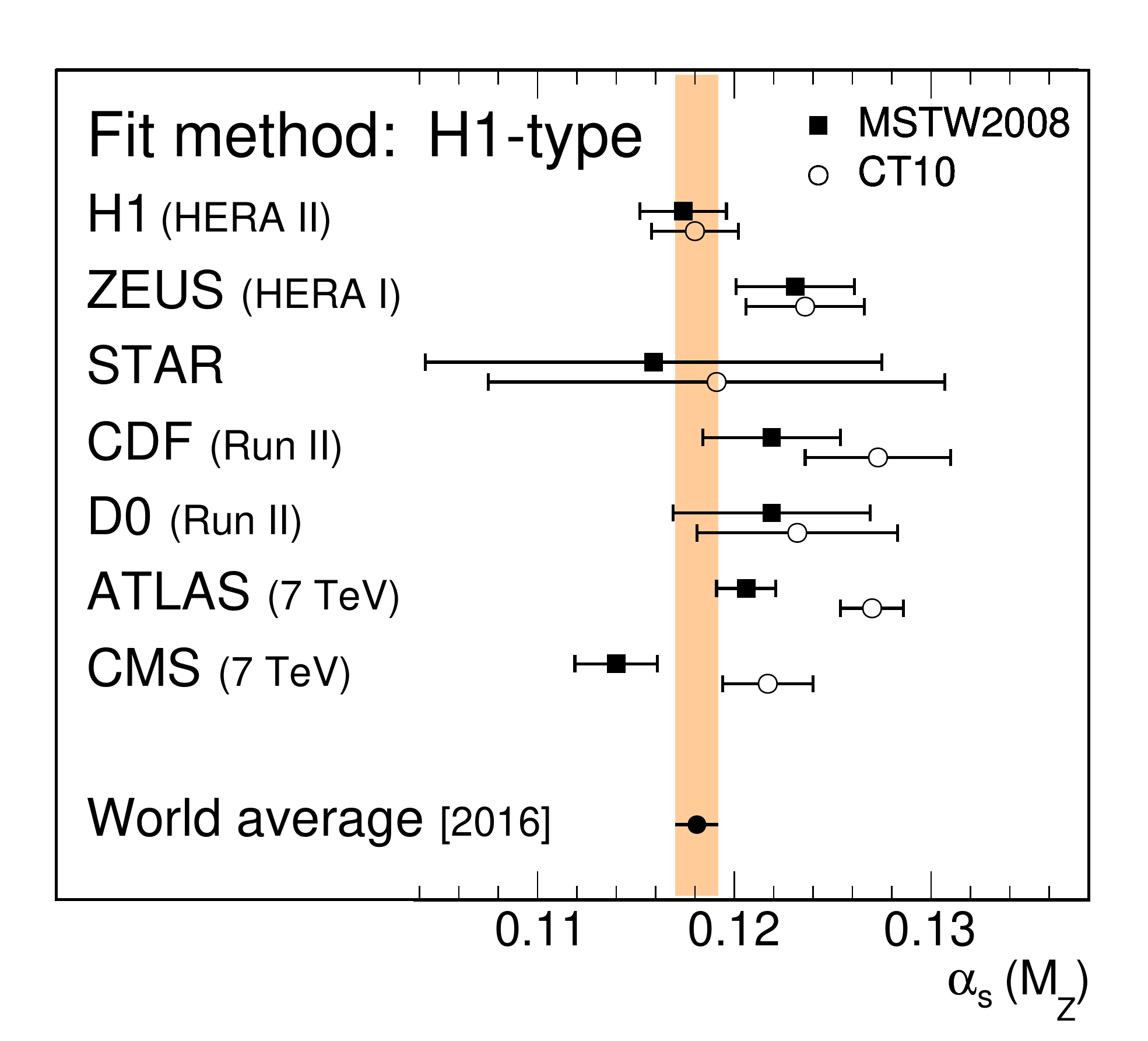}
  \end{center}
  \caption{Comparison of the \asmz results with their experimental
    uncertainties obtained using the H1-type extraction methods for
    CT10 and MSTW2008 PDFs. The world average
    value~\cite{Patrignani:2016xmw} is shown together with a band
    representing its uncertainty.}
  \label{fig:asvaluesPDF}
\end{figure}

The PDF dependence is displayed in figure~\ref{fig:asvaluesPDF}, where
the \asmz results for CT10 and MSTW2008 PDFs are compared to each
other, both obtained using the H1-type method.  While the PDF choice
has no significant effect for the results from the H1, ZEUS, and D0
data, smaller variations are seen for the CDF data, and a large
dependence for the ATLAS and CMS data.  Re-investigating this PDF
dependence in the context of a common determination of \asmz as
described in the next section, we observe that differences with
respect to the updated PDF sets, CT14~\cite{Dulat:2015mca} and
MMHT2014~\cite{Harland-Lang:2014zoa}, are reduced.

% -----------------------------------------------------------
% -----------------------------------------------------------
% -----------------------------------------------------------
\section{Determination of
  \texorpdfstring{\boldmath{\asmz}}{alpha\_s(M\_Z)} from multiple
  inclusive jet data sets}
\label{sec:common}

The analysis of multiple data sets requires their correlations to be
taken into account.  For the present study, measurements from
different colliders are considered to be uncorrelated because of the
largely complementary kinematic ranges of the data sets and different
detector calibration techniques.  Furthermore, investigations with
respect to H1 and ZEUS data~\cite{Abramowicz:2015mha}, CDF and D0
data~\cite{Tevatron:2014cka}, or ATLAS and CMS
data~\cite{CMS:2015mvl,CMS:2014yfa}, did not identify a relevant
source of experimental correlation.  This only leaves theoretical
uncertainties as a source of potential correlations in this study.
For the determination of NP effects and their uncertainties various
methods and MC event generators have been
employed~\cite{Andreev:2014wwa,Chekanov:2006xr,Abelev:2006uq,starnp,Abulencia:2007ez,Abazov:2008ae,Abazov:2009nc,Aad:2014vwa,Khachatryan:2014waa}.  While a consistent derivation of these
corrections with corresponding correlations is desirable, this is
beyond the scope of this analysis.  Hence, the NP correction factors
and their uncertainties are considered to be uncorrelated between the
different data sets.  In contrast, the PDF uncertainties and the
uncertainties due to the renormalisation and factorisation scale
variations are treated as fully correlated; the relative variations
with respect to the nominal scales are performed simultaneously for
all data sets.

The method employed for the simultaneous \as extraction combines
components of the individual methods outlined in the previous section
and is referred to as ``CMS-type method''.  The central \asmz result is
found in an iterative \chisq minimisation procedure adopted from the
H1-type method, where a normal distribution is assumed for the relative
uncertainties.  The exact \chisq formula is given by
equation~\eqref{eq:chisqCommon} of
appendix~\ref{appendix:chisqdefinition}.  Whereas in the H1-type \chisq
expression, only experimental uncertainties are taken into account,
the \our-type method also accounts for the NP and PDF uncertainties in
the \chisq expression, as in the CMS-type and D0-type methods.  This
\chisq definition treats variances as relative values and thus has
advantages, \eg when numerically inverting the covariance matrix.
Moreover, uncertainties of experimental and theoretical origin are put
on an equal footing.  As in the H1-type method, and in contrast to the
CMS-type and D0-type ones, only PDF sets obtained with a fixed value of
$\asmz=0.1180$ are employed in the determination of the central \asmz
result, leaving the \as dependence of the PDFs to be treated as a
separate uncertainty.

In summary, the individual contributions to the total uncertainty of
the \asmz result are evaluated as follows: The experimental
uncertainty (exp) is obtained from the Hesse
algorithm~\cite{James:1975dr} when performing the \asmz extraction
with only the uncertainties of the measurements included.  The NP and
PDF uncertainties are derived by repeating the \asmz extraction while
successively including the corresponding uncertainty contributions and
calculating the quadratic differences.  Further sources of systematic
effects are considered as follows:
\begin{compactitem}
\item The ``\pdfas'' uncertainty accounts for the initial assumption
  of $\asmzpdf=0.1180$ made in the PDF extraction, which is not
  necessarily consistent with the value of \asmz used in the pQCD
  calculation.  It is calculated as the maximal difference between any
  of the results obtained with PDF sets determined for $\asmzpdf =
  0.1170$, $0.1180$, and $0.1190$, and therefore covers a difference
  of $\Delta\asmzpdf=0.0020$, which is somewhat more conservative than
  the recommendation in ref.~\cite{Butterworth:2015oua}.
\item The ``\pdfset'' uncertainty covers differences due to the
  considered PDF set. These are caused by assumptions made on the data
  selection, parameterisation, parameter values, theoretical
  assumptions, or the analysis method for the PDF determination.
  It is defined as half of the width of the envelope of the results
  obtained with the PDF sets CT14~\cite{Dulat:2015mca},
  HERAPDF2.0~\cite{Abramowicz:2015mha},
  MMHT2014~\cite{Harland-Lang:2014zoa},
  NNPDF3.0~\cite{Ball:2014uwa} and ABMP16~\cite{Alekhin:2017kpj,Alekhin:2018pai}\@.
\item The uncertainty due to variations of the renormalisation and
  factorisation scales customarily is taken as an estimate for the
  error of a fixed-order calculation caused by the truncation of the
  perturbative series.
  It is obtained using six additional \asmz
  determinations, in which the nominal scales $(\mur,\muf)$ are varied
  by the conventional factors of $(1/2,1/2)$, $(1/2,1)$, $(1,1/2)$,
  $(1,2)$, $(2,1)$, and $(2,2)$.
  The scale factor combinations of $(1/2,2)$ and $(2,1/2)$ are customarily
  omitted~\cite{Catani:2003zt,Cacciari:2003fi,Banfi:2010xy}.
\end{compactitem}
The NP, PDF, \pdfas, \pdfset, and scale uncertainties are added in
quadrature to give the theoretical uncertainty (theo).  The total
uncertainty (tot) further includes the experimental uncertainty.

In the previous section, cf.\ figure~\ref{fig:asvalues}, it was found
that the \chisqndf values differ significantly from unity for some of
the data sets.  This necessitates to investigate in further detail the
consistency of the data within an individual data set as well as among
the different data sets.  Moreover, new PDF sets have become
available.  Therefore, the \our-type method is employed to extract \asmz
from each individual data set and for each of the PDFs ABMP16, CT14,
HERAPDF2.0, MMHT2014, and NNPDF3.0.  The resulting \chisqndf values
are displayed in figure~\ref{fig:chisqexclude} left.  Detailed
listings of the \asmz results and their uncertainties are given in
appendix~\ref{sec:commonresults}.

\begin{figure*}[tbp]
  \centering
  \includegraphics[width=0.50\textwidth]{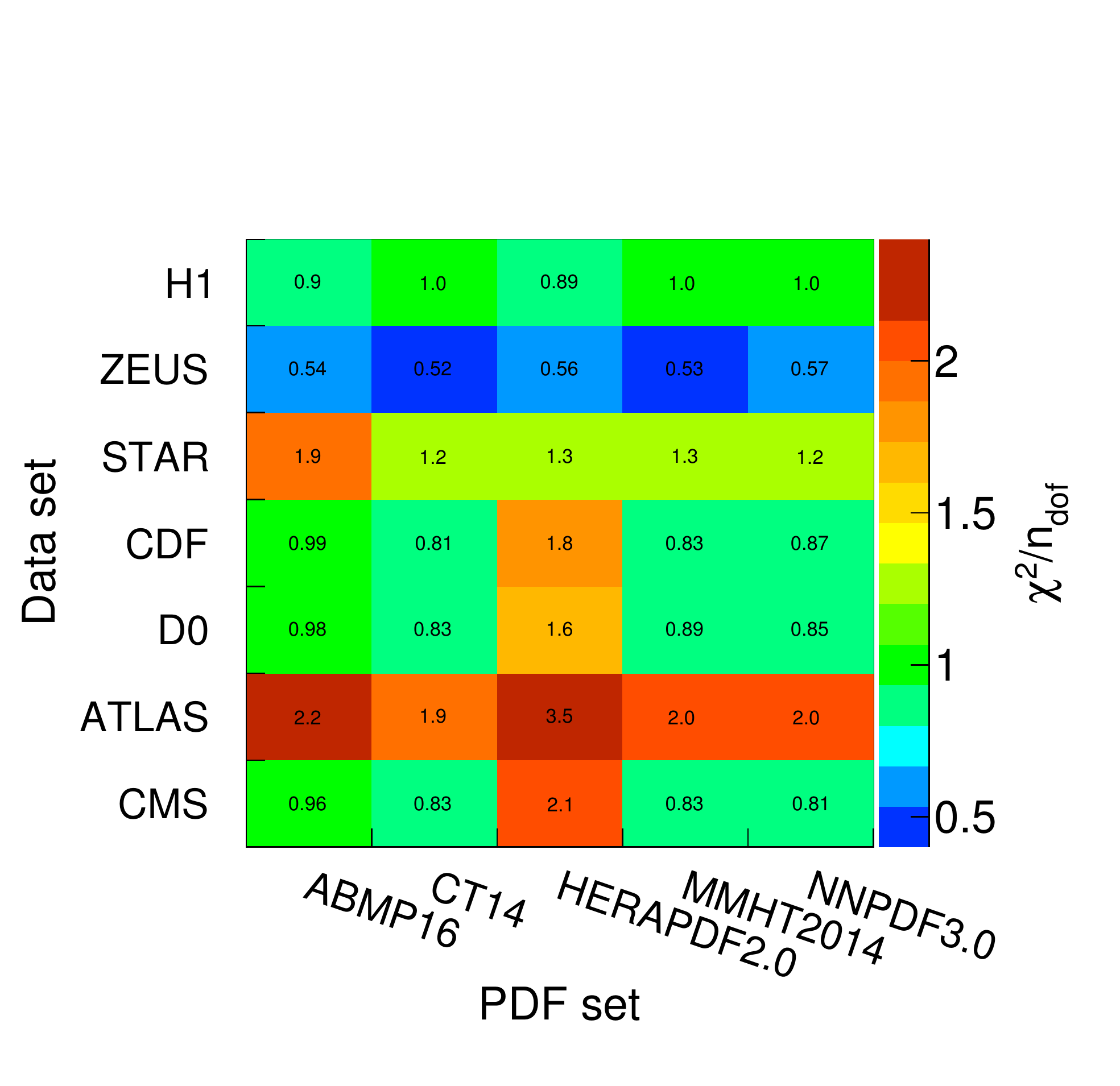}%
  \includegraphics[width=0.50\textwidth]{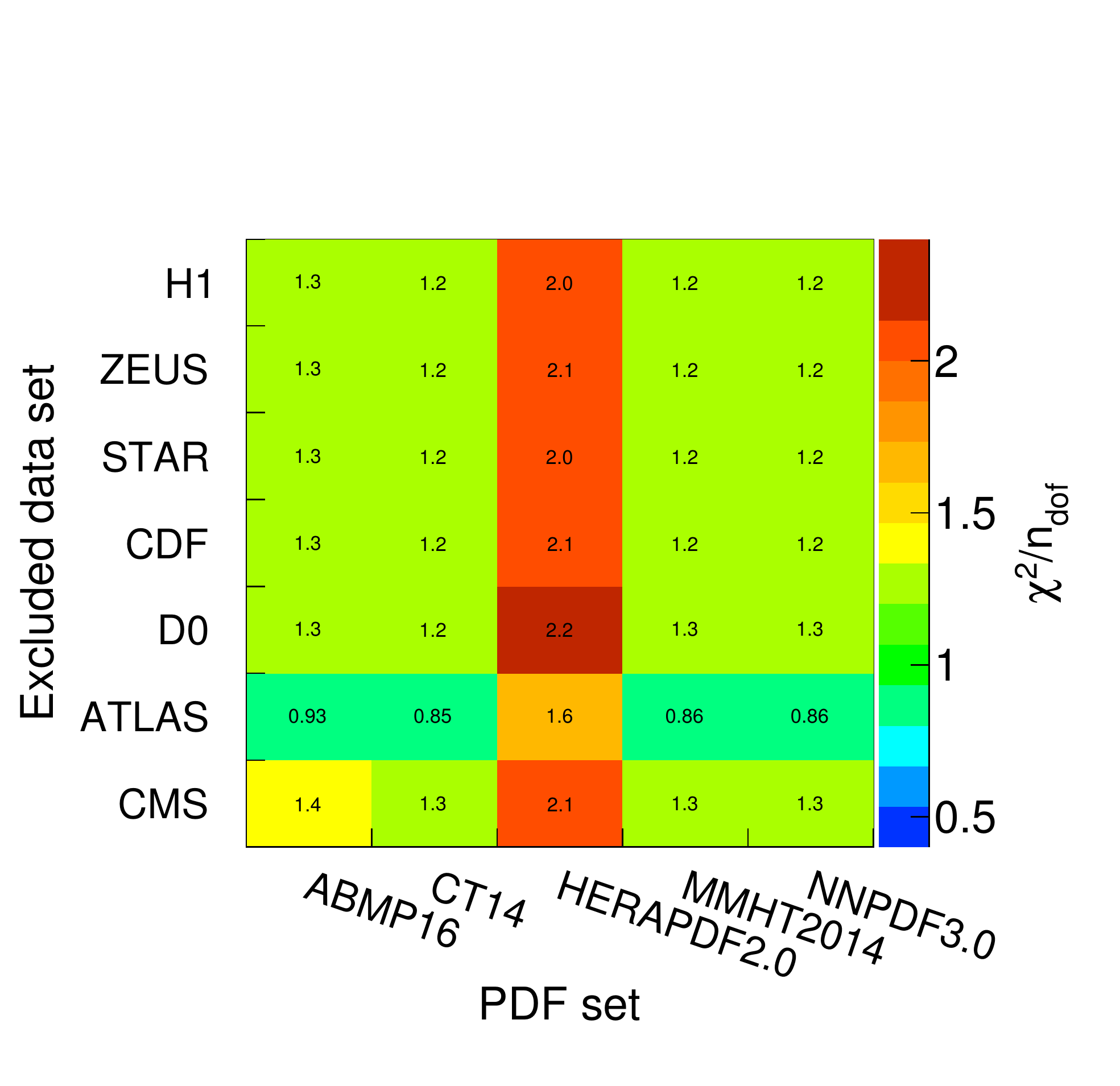}
  \caption{Left: Illustration of the \chisqndf values for fits to each
    data set individually. Right: Illustration of the \chisqndf values
    for simultaneous fits omitting a single data set at a time. The
    included or respectively excluded data set is indicated on the $y$
    axis and the PDF set on the $x$ axis. The fits are performed for
    each PDF set in the envelope definition of the \pdfset
    uncertainty.}
  \label{fig:chisqexclude}
\end{figure*}

For a given data set, \chisqndf is rather independent of the PDF set
used for the predictions and varies between 0.8 and 1.2.  These values
indicate reasonable agreement of the predictions with the data.
Exceptions are rather low values of \chisqndf around 0.54 found for
all PDF sets with the ZEUS data and large \chisqndf values between~1.9
and~3.5 exhibited by the ATLAS data, also for all PDFs. Exceptionally large
\chisqndf values appear for the \Tevatron or \LHC data together with
theory predictions using the HERAPDF2.0 set, and for the STAR data
in conjunction with the ABMP16 PDF set.

To further investigate the consistency among the data sets, a series
of \asmz extractions is performed, in which \asmz is determined
simultaneously from all data sets but one.  This is repeated for each
PDF set.  The resulting \chisqndf values are displayed in
figure~\ref{fig:chisqexclude} right.  Apparently, the exclusion of the
ATLAS data leads to significantly smaller \chisqndf values independent
of the PDF set used.  This hints at a compatibility issue when using
all data sets together, which is not present when the ATLAS data set
is ignored.  Therefore, we choose to exclude the ATLAS data for our
main result, which is thus obtained from the CDF, CMS, D0, H1, STAR,
and ZEUS inclusive jet data.  The choice of the NNPDF3.0 set for the
central result yields
\begin{eqnarray*}
  \asmz &=& 0.1192\,(12)_\text{exp}\,(5)_\text{NP}\\
  &&(7)_\text{PDF}\,(5)_\text{\pdfas}\,(11)_\text{PDFset}\,
  (^{+59}_{-38})_\text{scale}\,,
\end{eqnarray*}
\noindent{}with $\chisq = 328$ for 381 data points.  This result is
consistent with the world average value of
$0.1181\,(11)$~\cite{Patrignani:2016xmw}.  The experimental
uncertainty for the extraction from multiple data sets is
significantly smaller than each of the experimental uncertainties
reported previously for the separate \asmz determinations.  Results
obtained with different PDF sets constitute the \pdfset uncertainty.
They are listed in table~\ref{tab:asglobalPDF} together with the
PDF\footnote{For HERAPDF2.0, the PDF uncertainty does not include the
  ``model'' or ``parameterisation'' uncertainties as those are
  represented here by the \pdfset uncertainty.} and \pdfas
uncertainties as appropriate for the respective PDF set. The
corresponding values of \chisqndf can be read off from row six of
figure~\ref{fig:chisqexclude} right. %
Other uncertainties remain unchanged in the leading digit as compared
to the ones obtained for the NNPDF3.0 PDFs.

\begin{table*}[tbp]
%  \scriptsize
  \begin{center}
    \begin{tabular}{llrrrcrr@{\hskip44pt}r@{\hskip24pt}r}
      \toprule
      \multirow{2}{*}{\bf PDF set} %
      & \multirow{2}{*}{\bf\boldmath \asmz}
      & \multicolumn{8}{c}{{\bf Uncertainties} \footnotesize (scaled
        by a factor of $10^4$)}
      \\
      &
      & {exp}
      & {NP}
      & {PDF}
      & {\pdfas}
      & {\pdfset}
      & {scale}
      & {theo}
      & {total}
      \\\midrule
      ABMP16     & 0.1203 &           &          & $4$  & $3$ &
      &                & $^{+63}_{-47}$ & $^{+64}_{-49}$\\
      CT14       & 0.1206 &           &          & $10$ & $2$ &
      &                & $^{+58}_{-47}$ & $^{+59}_{-48}$\\
      HERAPDF2.0 & 0.1184 &           &          & $6$  & $2$ &
      &                & $^{+63}_{-51}$ & $^{+64}_{-53}$\\
      MMHT2014   & 0.1194 &           &          & $7$  & $3$ &
      &                & $^{+60}_{-46}$ & $^{+61}_{-48}$\\
      NNPDF3.0   & 0.1192 & $12$ & $5$ & $7$  & $5$ & $11$ &
      $^{+59}_{-38}$ & $^{+60}_{-41}$ & $^{+62}_{-43}$\\
      \bottomrule
    \end{tabular}
    \caption{Values of \asmz for the simultaneous fit to the H1, ZEUS,
      STAR, CDF, D0, and CMS data using the \our-type method for various
      PDF sets. The experimental, NP, \pdfset, and scale uncertainties
      remain mostly unchanged under a change of the PDF set and are
      quoted only once for NNPDF3.0.}
    \label{tab:asglobalPDF}
  \end{center}
\end{table*}

\begin{figure}[tbp]
  \begin{center}
    \includegraphics[width=0.54\textwidth]{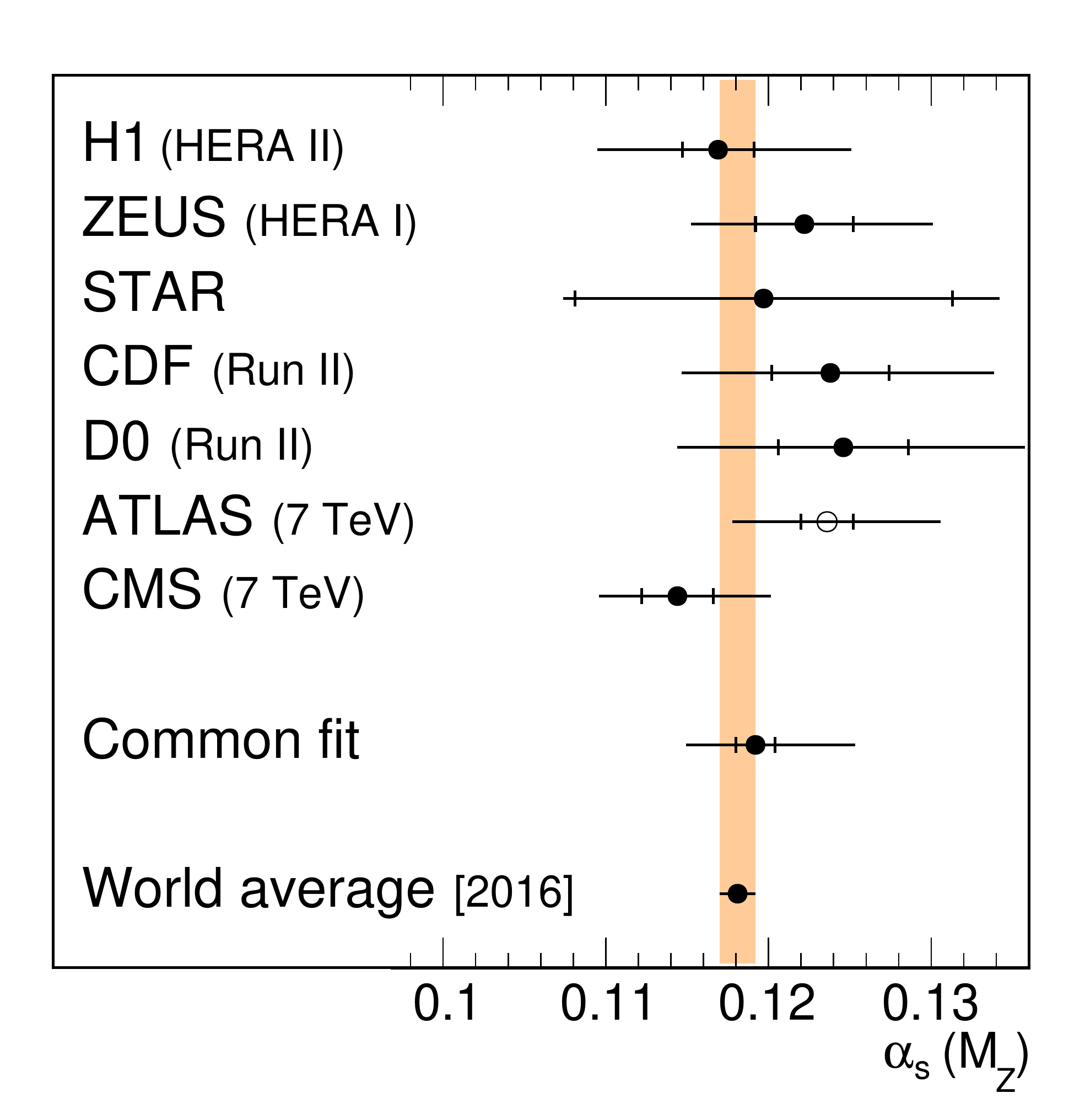}
  \end{center}
  \caption{The \asmz values from fits to individual data sets are
    compared to our simultaneous fit to H1, ZEUS, STAR, CDF, D0, and
    CMS data, and to the world average
    value~\cite{Patrignani:2016xmw}. The inner error bars represent
    the experimental uncertainty and the outer ones the total
    uncertainty. For reasons explained in the text, the ATLAS data are
    excluded from the common fit and only the result of a separate fit
    is indicated by the open circle.}
  \label{fig:common}
\end{figure}

The \asmz values from fits using the various PDF sets given in
table~\ref{tab:asglobalPDF} are found to be consistent within the
experimental uncertainty.  The NP, PDF, and \pdfas uncertainties are
smaller than the experimental uncertainty, while the \pdfset
uncertainty is of a similar size as the experimental one.  The scale
uncertainty is the largest individual uncertainty and is more than
three times larger than any other uncertainty.  Results of the \asmz
extractions from single data sets, cf.\
appendix~\ref{sec:commonresults}, from the simultaneous \asmz
extraction from all data sets, and the world
average~\cite{Patrignani:2016xmw} are compared in
figure~\ref{fig:common} and are seen to be consistent with each other.

\begin{figure*}[p]
  \centering
  \includegraphics[width=\textwidth]{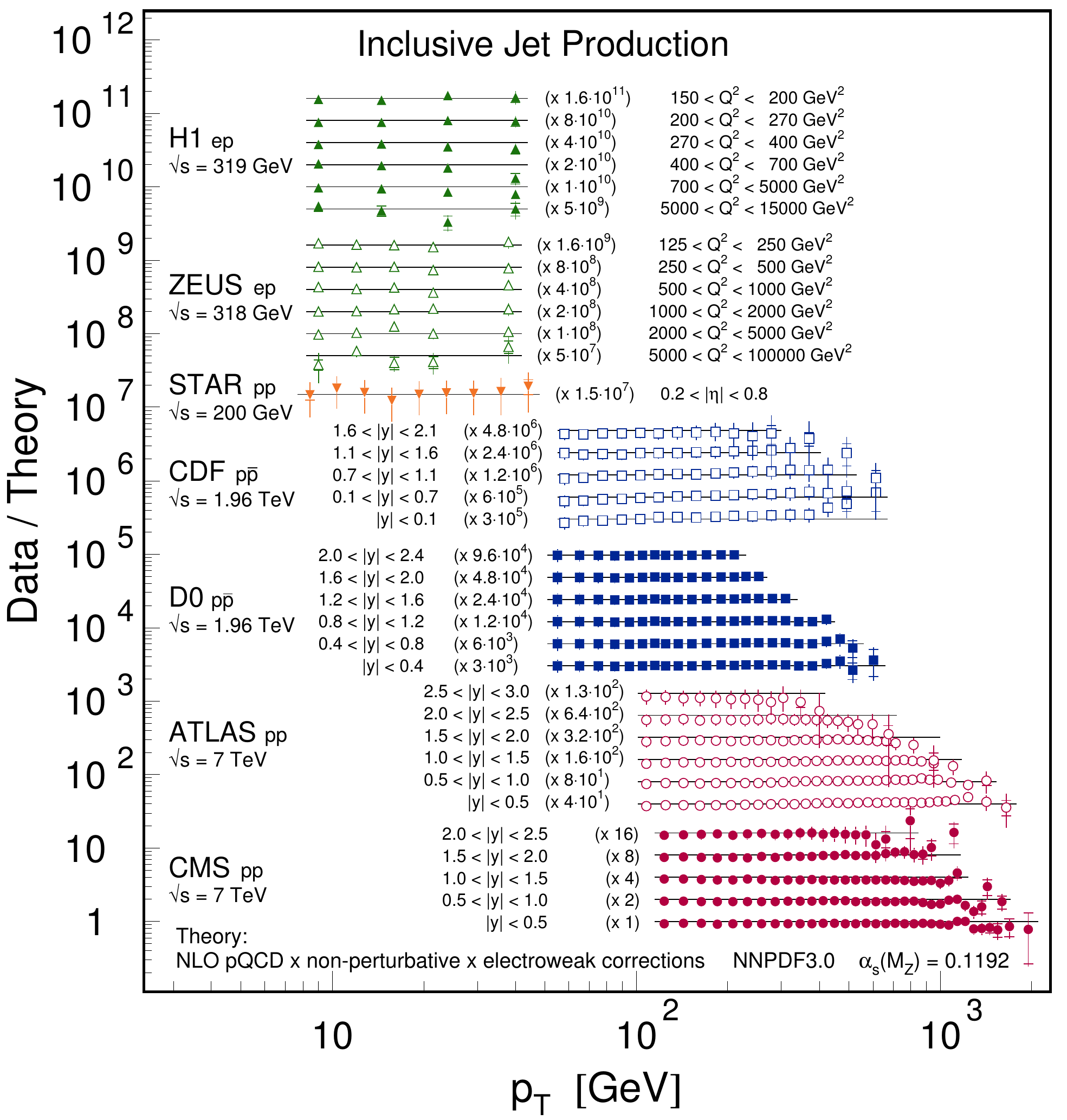}
  \caption{Ratio of data over theory for the selected inclusive jet
    cross sections listed in table~\ref{tab:datasets} as a function of
    jet \pt. The NLO predictions are computed with the NNPDF3.0 PDF
    set for the fitted \asmz value of $0.1192$, which is determined
    considering all presented data except for the ATLAS data set. They
    are complemented with non-perturbative corrections and, where
    appropriate, with electroweak corrections.}
  \label{fig:allincl}
\end{figure*}

The ratio of data to the predictions as a function of jet \pt for all
selected data sets is presented in figure~\ref{fig:allincl}.  The
predictions are computed for $\asmz=0.1192$ as obtained in this
analysis.  Visually, all data sets are well described by the theory
predictions.

% -------------------------------------------------------------------
% -------------------------------------------------------------------
% -------------------------------------------------------------------
%\clearpage
\section{Summary and outlook}
\label{sec:summary}

Inclusive jet cross section data from different experiments at various
particle colliders with jet transverse momenta ranging from 7\GeV up
to 2\TeV are explored for determinations of \asmz using
next-to-leading order predictions.

Previous \asmz determinations reported by the CMS, D0, and H1
collaborations~\cite{Khachatryan:2014waa,Abazov:2009nc,Andreev:2014wwa}
are taken as a baseline, and these \asmz extraction methods, which
differ in various aspects, are applied to inclusive jet cross section
data measured by the ATLAS, CDF, CMS, D0, H1, STAR, and ZEUS
experiments~\cite{Aad:2014vwa,Abulencia:2007ez,Khachatryan:2014waa,Abazov:2008ae,Andreev:2014wwa,Abelev:2006uq,Chekanov:2006xr}.
Differences\linebreak among the \asmz results due to the extraction technique
are found to be negligible in most cases. A new extraction method is
proposed, which combines aspects of the baseline approaches above.

In a statistical analysis, data measured by the CDF, CMS, D0, H1,
STAR, and ZEUS experiments are found to be well described by pQCD
predictions at next-to-leading order, and hence are considered to be
mutually consistent. Moreover, the values of \asmz determined from
each individual data set are found to be consistent among each
other. By determining \asmz simultaneously from these data, the
experimental uncertainty of \asmz is reduced to 1.0\,\%, as compared
to 1.9\,\% when only the single most precise data set of that
selection is considered.

The largest contribution to the uncertainty of \asmz originates from
the renormalisation scale dependence of the next-to-leading order pQCD
calculation. This uncertainty is expected to be reduced once the
next-to-next-to-leading order predictions become available for such
studies. Furthermore, a reevaluation of the non-perturbative
corrections and their uncertainties for all data sets in a consistent
manner is recommended for a determination of \asmz at high precision.
The presented study and the developed analysis framework provide a
solid basis for future determinations of \asmz and facilitate the
inclusion of additional data sets, further observables, and improved
theory predictions.

% -----------------------------------------------------------
% -----------------------------------------------------------
% -----------------------------------------------------------
%\acknowledgments%
\begin{acknowledgement}
We thank our colleagues in the CMS, D0, and H1 collaborations for
fruitful discussions, and K.~Bj{\o}rke and D.~Reichelt for early
related studies.
K.~Rabbertz thanks G.~Flouris, P.~Kokkas, and the colleagues from the
PROSA Collaboration.
D.~Savoiu acknowledges the support by the DFG-funded Doctoral School
``Karlsruhe School of Elementary and Astroparticle Physics: Science and
Technology''.
M.~Wobisch also wishes to thank the Louisiana Board of Regents Support
Fund for the support through the Eva J.~Cunningham Endowed
Professorship.
\end{acknowledgement}

% -----------------------------------------------------------
% -----------------------------------------------------------
% -----------------------------------------------------------
%\clearpage
\bibliography{brssw-pub}

% -----------------------------------------------------------
% -----------------------------------------------------------
% -----------------------------------------------------------
%\clearpage
\begin{appendix}

  % -------------------------------------------------------------------
  % -------------------------------------------------------------------
  % -------------------------------------------------------------------
  \section{Definition of the \texorpdfstring{\boldmath{\chisq}}{Chi2}
    expression for the \texorpdfstring{\our-type}{common-type} method}
  \label{appendix:chisqdefinition}

  In the \our-type method the \chisq expression, which is subject to the
  minimisation algorithm, is defined as
  \begin{multline}
    \chisq = %
    \sum_{ij}%
    \left(\log\frac{m_i}{t_i}\right)
    \left[\left(%
        \mathcal{V}_{\rm exp}+%
        \mathcal{V}_{\rm PDF}+%
        \mathcal{V}_{\rm NP}%
      \right)^{-1}\right]_{ij}\\
    \left(\log\frac{m_j}{t_j}\right)\,,
    \label{eq:chisqCommon}
  \end{multline}
  \noindent{}where the sum runs over all data points $i$ and $j$ of
  the measured cross sections $m_{i}$, $m_{j}$ and theory predictions
  $t_{i}$, $t_{j}$. The covariance matrices $\mathcal{V}$ represent
  the relative experimental, PDF, and NP uncertainties. A similar
  \chisq definition, taking into account only experimental
  uncertainties, was employed by the H1
  Collaboration~\cite{Andreev:2014wwa,Andreev:2016tgi,Andreev:2017vxu}.  For the calculation of the covariance matrices,
  all uncertainties are symmetrised, if necessary, by averaging the
  corresponding ``up'' and ``down'' shifts in \asmz in quadrature
  while keeping the sign of bin-to-bin correlations.  Uncertainty
  contributions to the total covariance matrix that are fully
  correlated across all observable bins in
  equation~\eqref{eq:chisqCommon} can alternatively be expressed in an
  equivalent form with nuisance parameters.

  % -------------------------------------------------------------------
  % -------------------------------------------------------------------
  % -------------------------------------------------------------------
  \section{Common-type extraction of
    \texorpdfstring{\boldmath{\asmz}}{alpha\_s(M\_Z)} from single
    inclusive jet data sets}
  \label{sec:commonresults}

  Detailed results of the \our-type method applied to the individual data
  sets are given in table~\ref{tab:ascommon}.  The result for the H1
  data agrees with the value published in ref.~\cite{Andreev:2014wwa}.
  Even though using the full D0 data set with 110 points, the
  extracted \asmz value is consistent with the value achieved by the
  D0 Collaboration at NLO for a subset of 22 points in
  ref.~\cite{Abazov:2009nc}.  For the CMS measurement, the \our-type
  method leads to a consistent but somewhat lower result than reported
  in ref.~\cite{Khachatryan:2014waa} for various PDFs.  Our result for
  the ZEUS data is compatible with the value obtained by the ZEUS
  Collaboration from a single-differential variant of the measurement
  in a reduced phase space as published in
  ref.~\cite{Chekanov:2006yc}.  With respect to the ATLAS, CDF, and
  STAR inclusive jet data, this study constitutes the first \asmz
  determination from either data set.  Within uncertainties, all \asmz
  values are consistent with each other and with the world average.

  \begin{table*}[tbp]
%    \scriptsize
    \begin{center}
      \begin{tabular}{llrrrrrr@{\hskip24pt}r@{\hskip24pt}r}
        \toprule
        \multirow{2}{*}{\bf Data set}  %
        & \multirow{2}{*}{\bf\boldmath \asmz}
        & \multicolumn{8}{c}{{\bf Uncertainties} \footnotesize (scaled by factor $10^4$)}
        \\
        &
        & {exp}
        & {NP}
        & {PDF}
        & {\pdfas}
        & {\pdfset}
        & {scale}
        & {theo}
        & {total}
        \\\midrule
        H1    %
        & $0.1169$
        & $22$ %
        & $9$ %
        & $8$ %
        & $4$ %
        & $10$ %
        & $^{+58}_{-47}$ %
        & $^{+60}_{-50}$ %
        & $^{+64}_{-54}$ %
        \\
        ZEUS  %
        & $0.1222$
        & $30$ %
        & $18$ %
        & $9$ %
        & $3$ %
        & $18$ %
        & $^{+48}_{-33}$ %
        & $^{+56}_{-42}$ %
        & $^{+63}_{-52}$ %
        \\
        STAR  %
        & $0.1197$
        & $116$ %
        & -- %
        & $50$ %
        & $26$ %
        & $99$ %
        & $^{+87}_{-41}$ %
        & $^{+143}_{-121}$ %
        & $^{+184}_{-168}$ %
        \\
        CDF  %
        & $0.1238$
        & $36$ %
        & $13$ %
        & $14$ %
        & $9$ %
        & $46$ %
        & $^{+83}_{-39}$ %
        & $^{+97}_{-64}$ %
        & $^{+104}_{-\phantom{0}73}$ %
        \\
        D0  %
        & $0.1246$
        & $40$ %
        & $23$ %
        & $21$ %
        & $8$ %
        & $62$ %
        & $^{+104}_{-\phantom{0}76}$ %
        & $^{+125}_{-103}$ %
        & $^{+131}_{-111}$ %
        \\
        ATLAS  %
        & $0.1236$
        & $16$ %
        & $3$ %
        & $15$ %
        & $8$ %
        & $30$ %
        & $^{+65}_{-34}$ %
        & $^{+74}_{-49}$ %
        & $^{+76}_{-51}$ %
        \\
        CMS  %
        & $0.1144$
        & $22$ %
        & $1$ %
        & $14$ %
        & $9$ %
        & $21$ %
        & $^{+58}_{-24}$ %
        & $^{+64}_{-36}$ %
        & $^{+68}_{-42}$ %
        \\\bottomrule
      \end{tabular}
      \caption{Results of \our-type \asmz extractions from individual
        inclusive jet data sets using the NNPDF3.0 PDF set. The values
        for \asmz are provided along with the experimental and
        theoretical uncertainties. The latter consist of contributions
        originating from NP effects, the propagation of the PDF
        uncertainties, the choices of the PDF \asmz value and the PDF
        set, and the scale uncertainty. The quadratic sum of the
        experimental and theoretical uncertainties is quoted as the
        total uncertainty. The corresponding \chisqndf values are
        displayed in column five of figure~\ref{fig:chisqexclude}
        left.}
      \label{tab:ascommon}
    \end{center}
  \end{table*}

  \noindent{}The individual uncertainties compare as follows:

  \begin{compactitem}
  \item The experimental uncertainty of \asmz is roughly comparable
    between experiments at the same collider.  It is largest for the
    STAR data, and smallest for the ATLAS data.
  \item The NP uncertainties are found to vary significantly, even
    between data sets in similar kinematic regions, for instance
    between CDF and D0\@.  In case of the \LHC experiments the NP
    uncertainties appear to be negligible.
  \item The PDF uncertainty as estimated with the NNPDF3.0 PDF set is
    smaller than the experimental uncertainty.  For the \HERA data,
    the PDF uncertainty is found to be moderately smaller than for
    \Tevatron or \LHC data as observed also with other PDF sets.
  \item For all data sets, the \pdfas uncertainty is rather small.
    This observation justifies to neglect the \as dependence of the
    PDFs in the \asmz determinations and to assign a separately
    derived uncertainty instead.
  \item The \pdfset uncertainty constitutes the largest contribution
    of the PDF related ones.
  \item The largely dominating scale uncertainty is of similar size in
    case of \HERA and \LHC data and somewhat larger for \Tevatron data
    or the STAR experiment.
  \end{compactitem}

  \noindent{}The results of \asmz determinations from single
  measurements for the alternative PDF sets ABMP16, CT14, HERAPDF2.0,
  and MMHT2014 PDF sets are provided in columns~2--5 of
  table~\ref{tab:asPDFvalues}. The envelope constructed from these
  four values together with the NNPDF3.0 result constitutes the
  \pdfset uncertainty shown in column seven of
  table~\ref{tab:ascommon}. The further columns in
  table~\ref{tab:asPDFvalues} present the PDF and \pdfas uncertainty
  for the respective PDF sets.

\begin{figure}[tbp]
    \begin{center}
      \includegraphics[width=0.45\textwidth]{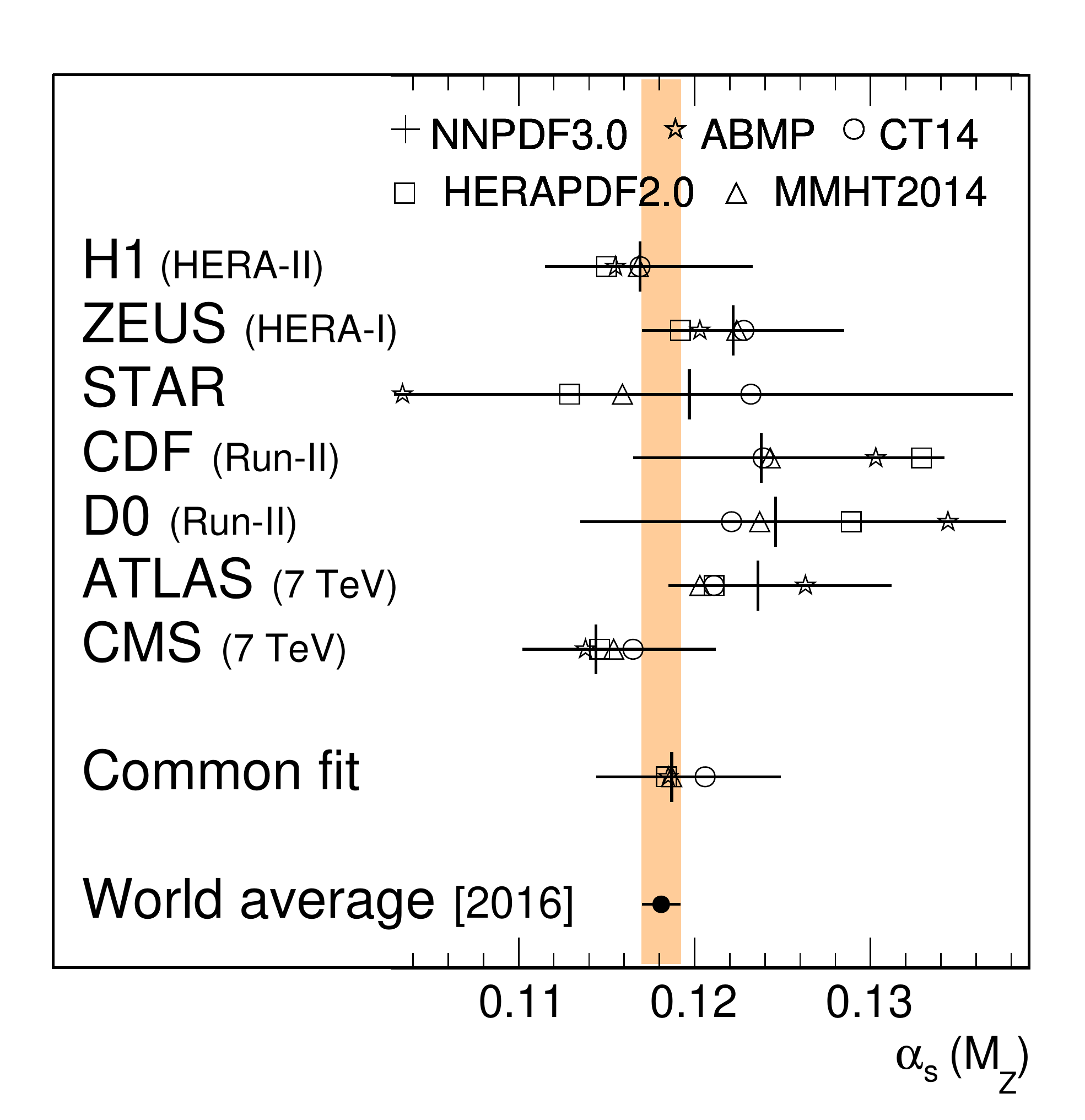}\hfill
    \end{center}
    \caption{Comparison of results for \asmz obtained with the
      alternative PDF sets ABMP16, CT14, HERAPDF2.0, and MMHT2014.
      The values are compared to the value of \asmz obtained with the
      NNPDF3.0 set and to the world average
      value~\cite{Patrignani:2016xmw}. The horizontal error bars,
      attached to the points representing the NNPDF3.0 results,
      indicate the total uncertainty.}
    \label{fig:PDFcomparison}
  \end{figure}

  The spread among the \asmz determinations from a single data set
  with varying PDF sets is illustrated in
  figure~\ref{fig:PDFcomparison}. For each of the individual data
  sets, the results are mostly consistent. Larger deviations are
  observed for the \Tevatron data when using the ABMP16 and HERAPDF2.0
  sets, and for the STAR data in conjunction with the\linebreak ABMP16~PDF set.

  The PDF uncertainty obtained with different PDF sets for the same
  data set is largest for CT14 and smallest for HERAPDF2.0. These
  numbers can differ by a factor of up to almost four. Moreover, we
  observe that in particular for \Tevatron and \LHC data the ABMP16
  and HERAPDF2.0 sets give significantly larger \pdfas uncertainties
  than the nominal PDF NNPDF3.0, whereas the CT14 or MMHT2014 PDFs
  exhibit in general systematically smaller \pdfas uncertainties than
  NNPDF3.0. A possible reason for the observed effects could lie in
  the different selections of data considered for the PDF
  determination. For instance, the ABMP16 and HERAPDF2.0 sets do not
  include any jet data.

  \begin{table*}[tbp]
%    \scriptsize
    \begin{center}
      \begin{tabular}{lcccc@{\hskip25pt}rrrr@{\hskip25pt}rrrr}
        \toprule
        \multirow{2}{*}{\bf Data set} &
        \multicolumn{4}{c}{\multirow{2}{*}{\bf\boldmath\asmz}} &
        \multicolumn{4}{l}{{\bf PDF uncertainty}} &
        \multicolumn{4}{l}{{\bf\boldmath\pdfas  uncertainty}}\\
        \multicolumn{5}{c}{} & \multicolumn{4}{l}{\footnotesize (scaled by factor $10^4$)} &
        \multicolumn{4}{l}{\footnotesize (scaled by factor $10^4$)}\\
              & AB & CT & H2 & MM & AB & CT & H2 & MM & AB & CT & H2 & MM\\\midrule
        H1    & 0.1155 & 0.1169 & 0.1150 & 0.1168 &
                4 & 11 &  3 &  7 &  9 &  3  &  8  &  4\\
        ZEUS  & 0.1203 & 0.1228 & 0.1192 & 0.1224 &
                5 & 11 &  4 &  7 &  9 &  1  &  8  &  2\\
        STAR  & 0.1034 & 0.1232 & 0.1129 & 0.1159 &
               22 & 63 & 30 & 37 & 18 &  5  &  1  & 12\\
        CDF   & 0.1303 & 0.1239 & 0.1329 & 0.1243 &
               13 & 29 &  8 & 19 & 27 &  1  & 17  &  1\\
        D0    & 0.1344 & 0.1221 & 0.1289 & 0.1237 &
               19 & 34 & 16 & 23 & 29 &  2  & 11  &  0\\
        ATLAS & 0.1263 & 0.1211 & 0.1211 & 0.1203 &
               13 & 22 & 13 & 17 & 11 &  0  &  6  &  3\\
        CMS   & 0.1186 & 0.1165 & 0.1146 & 0.1154 &
               10 & 24 & 14 & 19 & 14 &  2  &  6  &  2\\\bottomrule
      \end{tabular}
      \caption{Results of \our-type fits to single inclusive jet data
        sets for varying PDF sets. Listed are the \asmz results and
        the respective PDF and \pdfas uncertainty. For the purpose of
        a more compact presentation, the employed PDF sets ABMP16
        (AB), CT14 (CT), HERAPDF2.0 (H2), and MMHT2014 (MM) are
        abbreviated here to the two-letter acronyms given in
        parentheses.  Other uncertainty components differ only in the
        last digit as compared to the results obtained with NNPDF3.0
        displayed in table~\ref{tab:ascommon}. Uncertainties are
        scaled by a factor of $10^4$.}
      \label{tab:asPDFvalues}
    \end{center}
  \end{table*}

  %% ------------------------------------------------------------------------ %%
\end{appendix}

%% ------------------------------------------------------------------------ %%

\end{document}